\documentclass[11pt]{article}
\usepackage[pctex32]{graphics}
\usepackage{amsmath,amsthm}
\usepackage{amssymb,latexsym}
\usepackage[mathscr]{eucal}
\usepackage{setspace}

\newcommand{\bee}{\begin{equation}}
\newcommand{\ene}{\end{equation}}

\newcommand{\p}{^\prime}

\begin{document}

\vspace*{.8in}

\begin{center}
{\bf Kuzmin-Oseledets Formulations of \\Compressible Euler Equations}

\vspace{.5in}
Bhimsen Shivamoggi\footnote{Permanent Address: University of Central Florida, Orlando, Florida, 32816-1364}, Susan Kurien and Daniel Livescu\\
Los Alamos National Laboratory\\
Los Alamos, NM 87545\end{center}

\vspace{.5in}

\noindent
{\bf Abstract}

\doublespace

Kuzmin-Oseledets formulations of compressible Euler equations case are considered. Exact results and physical interpretations are given. One such exact result for the compressible barotropic case is the potential helicity Lagrange invariant. In recognition of the fundamental physical implications of this Lagrange invariant clarified here, this invariant is shown to hold for compressible non-barotropic cases as well upon using a stronger gauge condition. Symmetry restoration taking place at the Lagrangian level in the Kuzmin-Oseledets formulation is pointed out. The Kuzmin-Oseledets formulation in the compressible barotropic case is shown to admit an \emph{exact} solution that physically describes a density wave on a steady irrotational flow with the Kuzmin-Oseledets velocity \textbf{\emph{q}} growing monotonically with it and hence specifying some information about the fluid impulse that is needed to set up the flow in question. 

\vspace{4.0in}

\pagebreak

\noindent\large\textbf{1 Introduction}

\normalsize
Impulse formulations of Euler (and Navier-Stokes) equations were considered by Kuzmin [1] and Oseledets [2]. Different impulse formulations are produced by various possible gauge transformations (Russo and Smereka \cite{Rus}). In the Kuzmin-Oseledets gauge, the impulse variable {\bf q} has an interesting geometrical meaning: it describes the evolution of material surfaces; its direction is orthogonal to the material surface element, and its length is proportional to the area of the surface element. The extension of the Kuzmin-Oseledets formulation to the compressible barotropic case was considered in a brief way by Tur and Yanovsky \cite{Tur}. In this paper, we give first a reformulation of this aspect which we then use to develop further exact results and physical interpretations. One such exact result for the compressible barotropic case is the potential helicity Lagrange invariant. In recognition of the fundamental physical implications of this Lagrange invariant clarified here, this invariant is shown to hold for compressible non-barotropic cases as well upon using a stronger gauge condition. Another exact result is an \emph{exact} solution that physically describes a density wave on a steady irrotational flow with the Kuzmin-Oseledets velocity \textbf{\emph{q}} growing monotonically with it and hence specifying some information about the fluid impulse that is needed to set up the flow in question. 

\bigskip

\noindent\large\textbf{2 Kuzmin-Oseledets Formulations of Compressible Euler Equations}

\normalsize
We give below the Kuzmin-Oseledets type formulation of compressible Euler equations as a preamble to the exact results given in Sections 3 and 4. This is conducive to the development of the latter results. By contrast, the previous brief treatement of this aspect by Tur and Yanovsky \cite{Tur} proceeded directly with a Lie-group analysis of equations (1) and (2) (below). Euler equations for a compressible fluid are
\bee
\frac{\partial \rho}{\partial t} + {\bf \nabla} \cdot (\rho {\bf v}) = 0
\ene
and
\bee
\frac{\partial {\bf v}}{\partial t} + ({\bf v} \cdot {\bf \nabla}) {\bf v} = -\frac{1}{\rho} {\bf \nabla} p.
\ene

\noindent
For a barotropic case, namely,
\bee
p = p (\rho)
\ene
equation (2) may be rewritten as
\bee
\frac{\partial {\bf v}}{\partial t} - {\bf v} \times ({\bf \nabla} \times {\bf v}) = - {\bf \nabla}(P + \frac12 {\bf v}^2)
\ene
where
\bee
P(\rho) \equiv \int\frac{dp}{\rho}.
\ene

Introduce the Helmholtz decomposition -
\bee
{\bf q} = {\bf v} + {\bf \nabla} \phi
\ene
$\phi$ being an arbitrary scalar field; {\bf q} then evolves, from equation (4), according to
\bee
\frac{\partial {\bf q}}{\partial t} - {\bf v} \times ({\bf \nabla} \times {\bf q}) = - {\bf \nabla} (P + \frac12 {\bf v}^2 - \frac{\partial \phi}{\partial t})
\ene

Imposing the following gauge condition on $\phi$ :
\bee
\frac{\partial \phi}{\partial t} + ({\bf v} \cdot {\bf \nabla})\phi + \frac12 {\bf v}^2 - P = 0.
\ene

\noindent
equation (7) becomes
\bee
\frac{\partial {\bf q}}{\partial t} + ({\bf v} \cdot {\bf \nabla}) {\bf q} = - ({\bf \nabla} {\bf v})^T {\bf q}.
\ene

On the other hand, taking the curl of equation (4) or (7) and using equation (1), we obtain the following equation for the potential vorticity $\omega$/$\rho$ - 
\bee
\frac{\partial}{\partial t} (\frac{\omega}{\rho}) + ({\bf v} \cdot {\bf \nabla})(\frac{\omega}{\rho}) = (\frac{\omega}{\rho} \cdot {\bf \nabla}) {\bf v}.
\ene

where,

\bee
{\bf \omega}\equiv \nabla \times {\bf V} = \nabla \times {\bf q}.
\ene

\pagebreak
\noindent\large\textbf{3 General Results}

An immediate implication of (11) is that vorticity flux conservation holds the Kuzmin-Oseledets velocity {\bf q} as well. This is set forth by the following theorem.

\noindent\normalsize
{\bf Theorem 1:} The Kuzmin-Oseledets velocity {\bf q}, as defined in (6), satisfies for a compressible barotropic fluid the Kelvin-Helmholtz circulation theorem 
\bee
\frac{d}{dt}\oint_C {\bf q} \cdot {\bf dl} = 0
\ene
where C is a closed material curve in the fluid.

\noindent
{\it Proof}: We have for a compressible barotropic fluid (Batchelor \cite{Bat}),
\bee
\frac{d}{dt}\oint_C {\bf v} \cdot {\bf dl} = 0
\ene
which, on using (6), leads to
\bee
\frac{d}{dt}\oint_C {\bf q} \cdot {\bf dl} = 0\quad or\quad
\oint_C {\bf q} \cdot {\bf dl} = const.
\ene

On the other hand, the simplified equation (9) for the evolution of the Kuzmin-Oseledets velocity {\bf q} in conjunction with the potential vorticity evolution equation (10) leads to Lagrangian invariance of the potential helicity $ ({\bf q} \cdot{\bf \omega}/\rho)$.

\noindent
{\bf Theorem 2:} The compressible barotropic flow has the potential helicity Lagrange invariant -
\bee
[\frac{\partial}{\partial t} + ({\bf v} \cdot {\bf \nabla})](\frac{{\bf q} \cdot {\bf \omega}}{\rho}) = 0.
\ene

\noindent
{\it Proof}: (15) follows immediately from equations (9) and (10).

(15) was given by Tur and Yanovsky \cite{Tur} who used a Lie group approach to equations (1) and (2), but the physical interpretation of (15) was not recognized which we develop in the following.

If {\bf l} is a vector field associated with an infinitesimal line element of the fluid, {\bf l} evolves according to (Batchelor \cite{Bat}) -
\bee
[\frac{\partial}{\partial t} + ({\bf v} \cdot {\bf \nabla})] {\bf l} = ({\bf l} \cdot {\bf \nabla}) {\bf v}
\ene
which is identical to the potential vorticity equation (10). Therefore, the potential vortex lines evolve as fluid line elements.

On the other hand, if {\bf S} is a vector field associated with an oriented material surface element of the fluid, {\bf S} evolves according to (Batchelor \cite{Bat})-
\bee
[\frac{\partial}{\partial t} + ({\bf v} \cdot {\bf \nabla})](\rho{\bf S}) = -({\bf \nabla} {\bf v})^T (\rho{\bf S})
\ene
which is identical to the equation of evolution of the Kuzmin-Oseledets velocity {\bf q}, namely, equation (9). Therefore, the field lines of {\bf q} evolve as fluid mass surface elements.

Thus, the potential helicity Lagrange invariant
\bee
\frac{{\bf q} \cdot {\bf \omega}}{\rho} = const
\ene
is simply physically equivalent to the mass conservation of the fluid element.

It is of interest to note that in the Eulerian picture the usual helicity integral

\bee
H \equiv \int_V {\bf v} \cdot {\bf \omega} dV
\ene

\noindent
where V is a material volume in the fluid, continues to be an invariant even in the compressible barotropic case (Moffatt \cite{Mof}) which is perhaps plausible because of the lack of any kinematic connection for (19) unlike (18).

On the other hand, the Lagrange invariant (18) also implies
\bee
{\bf q} \cdot {\bf l} = const
\ene
which may be seen to be a sufficient condition for the validity of the circulation conservation (12) for the Kuzmin-Oseledets velocity {\bf q}.

Thus, the conservation laws of mass and momentum (and hence kinematics and dynamics) apparently undergo a certain merger at the Lagrangian level in the Kuzmin-Oseledets formulation signifying some symmetry restoration taking place there!

It is of interest to note that the potential helicity Lagrange invariant (18) offers a new physical perspective on the Beltrami state for a compressible barotropic fluid (Shivamoggi and van Heijst\cite{Shi}) given by

\bee
{\bf \omega} = a\rho{\bf v}.
\ene

Using (21), (18) leads to 

\bee
{\bf q}\cdot {\bf v} = const\ or\  (\rho {\bf S})\cdot {\bf v}=const
\ene

\noindent
which signifies the constancy of mass flux that is to be expected since the Beltrami state (21) is steady!

\noindent\large\textbf{4 An Exact Solution}

\normalsize
The Kuzmin-Oseledets formulation in the compressible barotropic case admits an \emph{exact} solution. In order to obtain this, consider the velocity field in $(r, \theta, z)$ coordinates -

\bee
{\bf v} = <V, \frac{\zeta}{r}U(\zeta), 0>
\ene
where $V$ is a constant that characterizes the non-solenoidal aspects of the velocity {\bf v} and hence may be viewed as the compressibility parameter, the incompressibility limit being given by $ V \rightarrow 0 $.
\bee
\zeta \equiv r-V t.
\ene

The vorticity associated with (21) is
\bee
\omega = <0, 0, \frac{1}{r}[\zeta U(\zeta)]'>.
\ene

Using (23), the mass conservation equation (1) leads to
\bee
\frac{\partial}{\partial t}(r\rho) + V\frac{\partial}{\partial r}(r\rho) = 0
\ene
from which
\bee
\rho = \frac{1}{r}g(\zeta).
\ene

The incompressible limit condition
\bee
V = 0\quad:\quad \rho=const
\ene
on application to (27), leads to
\bee
g(\zeta) = a\zeta
\ene
$a$ being an arbitrary constant.

Using (29), (27) leads to
\bee
\rho = \frac{a}{r}\zeta.
\ene
Here, the parameters $a$ and V have to be chosen appropriately so as to keep $\rho$ positive definite.

Next, equation (9) leads to
\bee
\frac{\partial q_r}{\partial t} + V \frac{\partial q_r}{\partial r} - (\frac{\zeta}{r}U)\frac{q_\theta}{r} = - \frac{q_\theta}{r}\frac{\partial}{\partial r}(\zeta U)-\frac{q_r V}{r}
\ene

\bee
\frac{\partial}{\partial t} (rq_\theta) + V \frac{\partial}{\partial r} (rq_\theta) = 0.
\ene

Equation (32) yields,
\bee
rq_\theta = f(\zeta).
\ene

Now, the equi-vorticity condition associated with (6), namely,
\bee
{\bf \nabla} \times q = {\bf \nabla} \times {\bf v}
\ene
gives,
\bee
q_\theta = \frac{\zeta}{r}U(\zeta).
\ene

Comparing (35) with (33), we have
\bee
f(\zeta) = \zeta U(\zeta).
\ene

Using (33) and (36), equation (31) becomes
\bee
\frac{\partial}{\partial t} (rq_r) + V\frac{\partial}{\partial r}(rq_r) = - \frac{\zeta^2}{r}UU' - \frac{\zeta V t}{r^2}U^2.
\ene

On putting,
\bee
rq_r = \frac{\zeta^2}{r}G(\zeta)t
\ene
equation (37) leads to
\bee
\zeta G - \frac{\zeta V t}{r}G = -\zeta UU'-\frac{V t}{r}U^2
\ene
from which, for the compressible case $(V\neq 0),$ we obtain

\begin{equation}
G = \frac{U^2}{\zeta}
\end{equation}

\noindent
and hence,

\bee
\zeta U' + U = 0.
\ene

It may be noted, on the other hand, that on substituting (23) and (25), equation (10) yields 

\begin{subequations}
\bee
\frac{\partial}{\partial t} (\zeta U\p + U) + rV\frac{\partial}{\partial r}[\frac{1}{r}(\zeta U\p +U)] =0
\ene

\noindent
or

\bee
-\frac{V}{r}(\zeta U\p +U) = 0
\ene
\end{subequations}

\noindent
which, for the compressible case $(V\neq 0)$, leads again to equation (41)

\indent
Equation (41) yields -

\bee
U(\zeta) = \frac{b}{\zeta}
\ene

\noindent
$b$ being an arbitrary constant. Using (43), (40) yields

\bee
G(\zeta) = \frac{b^2}{\zeta^3}
\ene

Using (43) and (44), we obtain from (35) and (38) -

\bee
q_r = \frac{b^2t}{r^2\zeta}\quad,\quad q_\theta = \frac{b}{r},
\ene
while (23) becomes
\bee
{\bf v} = <V, \frac{b}{r}, 0>.
\ene

(41) or (46) implies that this flow has zero vorticity -

\bee
{\bf \omega} ={\bf \nabla}\times {\bf v} = {\bf \nabla}\times {\bf q} = {\bf 0}.
\ene

Thus, the flow under consideration is, as implied by (30), (45), and (46), a density wave on a steady irrotational flow with the corresponding Kuzmin-Oseledets velocity {\bf q} growing monotonically with $t$. The latter aspect provides some information about the fluid impulse that is needed to set up the flow in question.

\bigskip

\noindent\large\textbf{5 Generalization to Non-barotropic Cases}

The fundamental physical nature of the Lagrange invariant (18) suggests its validity beyond the barotropic constraint that was invoked in its derivation in Section 3. However, a gauge condition stronger than (8) becomes now necessary to deduce this result.

For the non-barotropic case, equations (9) and (10) become

\bee
\frac{\partial {\bf{q}}}{\partial t}+(\textbf{v}\cdot\nabla)\textbf{v} = -(\nabla\textbf{v})^{T}\textbf{q} - \frac{1}{\rho}\nabla p + \nabla \left[ \frac{\partial\phi}{\partial t} + (\textbf{v}\cdot\nabla)\phi + \frac{1}{2}\textbf{v}^{2}
\right]
\ene

\bee
\frac{\partial}{\partial t} 
\left( \frac{{\bf{\omega}}}{\rho} \right) + ({\bf{v}} \cdot \nabla) \left(\frac{\omega}{\rho}\right) = \left( \frac {\bf\omega}{\rho}\cdot\nabla \right){\bf{v}} - \frac{1}{\rho}\nabla\times\left( \frac{1}{\rho}\nabla p\right). 
\ene

Using equations (48) and (49), we may derive the following equation - 

\bee
\left[ \frac{\partial}{\partial t} + ({\bf{v}}\cdot\nabla)\right] \left( \frac { {\bf{q}}\cdot{\bf{\omega}}}{\rho} \right) = - \frac{1}{\rho}\left[{\bf{q}} \cdot \nabla \times + ( \nabla \times {\textbf{q}}).\right] \left[ \frac{1}{\rho}\nabla p - \nabla \left\{ \frac{\partial \phi}{\partial t} + ({\bf{v}} \cdot\nabla)\phi + \frac{1}{2}{\bf{v}}^{2}\right\} \right]. 
\ene

\textbf{Theorem 3:} The compressible non-barotropic flow has the potential helicity Lagrange invariant - 

\bee
\left[ \frac{\partial}{\partial t} + ({\bf{v}}\cdot\nabla)\right] \left(\frac{{\bf{q}}\cdot{\bf{\omega}}}{\rho}\right) = 0. 
\ene

\emph{Proof}: (51) follows immediately from equation (50) on imposing the following stronger gauge condition on $\phi$ - 

\bee
\left[ {\bf{q}}\cdot\nabla \times + (\nabla\times {\textbf{q}}) \cdot\right] \left[ \frac{1}{\rho}\nabla p - \nabla\left\{ \frac{\partial\phi}{\partial t} + ({\bf{v}}\cdot\nabla)\phi + \frac{1}{2} {\bf{v}}^{2}\right\}\right] = 0. 
\ene

\bigskip

\noindent\large\textbf{6 Discussion}

\normalsize\noindent
In this paper, Kuzmin-Oseledets formulations of compressible Euler equations for the barotropic case are considered. For the compressible barotropic case, the potential helicity Lagrange invariant and its physical interpretation are given. In recognition of the fundamental physical implications of this Lagrange invariant clarified here, this invariant is shown to hold for compressible non-barotropic cases as well upon using a stronger gauge condition. The kinematics and dynamics aspects are indicated to apparently undergo a certain unification at the Lagrangian level in the Kuzmin-Oseledets formulation. These symmetries seem to break as one moves up to the Eulerian level. The Kuzmin-Oseledets formulation in the compressible barotropic case is shown to admit an \emph{exact} solution that physically describes a density wave on a steady irrotational flow with the Kuzmin-Oseledets velocity {\bf q} growing monotonically with $t$. The latter aspect provides some information about the fluid impulse that is needed to set up the flow in question.

\pagebreak

\end{document}